# *Hybrid* Models of Step Bunching

Diana Staneva, Bogdan Ranguelov, Vesselin Tonchev,

Institute of Physical Chemistry, Bulgarian Academy of Sciences, 1113 Sofia, Bulgaria

**Abstract.** We introduce two *hybrid* models of step bunching on vicinal crystal surfaces. The model equations for step velocity are constructed by the two possible exchanges of terms between the equations of two *primary* models MM2 and LW2 [arXiv:1011.1863], both showing the specific type of bunching with minimal step-step distance $l_{min}$ in the bunch independent of the number of steps $N$ in it. This feature is preserved only in the hybrid model LW2MM (the first term in the model equation comes from LW2 and the second one - from MM2) but in a rather complex fashion – the surface slope is largest in the both ends of the bunch and after a sharp decrease jumps again to become constant in the inner part. We restrict our considerations to the simplest case of $p = 0$, $p$ being the exponent in the destabilizing term in the velocity equations. The time-scaling exponent of $N$ in LW2MM is ~1/3 and is independent of $n$, the exponent in the stabilizing term of the velocity equations. The other model, MM2LW, shows an interesting type of step bunching – some bunches grow to a certain size and then decay emitting steps towards the two adjacent bunches. The bunch compression with the increase of $N$ *is* pronounced.

**Introduction.** Vicinal crystal surfaces are a result of the discreteness of matter – only few high symmetric (low-index) crystal surfaces are free of steps while the arbitrary cuts result in regular stairways on the atomic length scale. The steps are originally with unit height – if the building units of the crystal are atoms, the steps are called mono-atomic. The vicinal surfaces are focus of significant research efforts since the technologically important growth mode, the so called *step flow growth*, is realized through the attachment of single building units to the steps. Thus it is important to preserve the equidistant step arrangement and a phenomenon with an opposite effect is the *step bunching* – due to various reasons of both kinetic and thermodynamic nature the steps could gather in bunches leaving large areas between the bunches free of steps. Recently the bunched surfaces are becoming subject of technological importance of its own as nano-templates for different growth strategies [1] since bunch formation makes the crystal surface functionally and structurally inhomogeneous. The interest in studying step bunching on vicinal crystal surfaces was boosted by the observation of the phenomenon on Si(111)-vicinals [2]. Since then numerous studies were reported in the three main research directions – experimental, theoretical and numerical, and remarkable exchange and mutual challenges took place. So far, two rather general types of step bunching could be identified according to the behavior of the average (minimal) inter-step distance in the bunch $l_b$ ($l_{min}$) with the increase of the number of steps *N* in it - B2-type in which both distances decrease with *N* [3] and B1-type with distances independent of *N* [4, 6]. In other words in B2-type the surface slope inside the bunches increases with *N* while in the B1-type the slope is independent of *N*. The number in the notation of the type, 1 or 2, reflects the number of characteristic length scales necessary to describe the step bunches in either of the types. Thus, to describe the step bunching in B2-type one needs both the width and the height of the bunch while in B1-type there is a linear dependence between these. The 'classical' mechanisms of surface destabilization, including electromigration and Ehrlich-Schwoebel effect, result in B2-type, while the effective step-step attraction, emerging usually in strained hetero-epitaxial layers, results in B1-type step bunching. The quest for new models of various types, being one of the signs for the maturity of the field, is dictated by several reasons besides the main one – to explain and predict better the experimental observations. Another important stimulus is the development of general picture of step bunching phenomena in terms of universality classes [5, 8] which approach is originally developed considering a generalized continuum equation for the time evolution of the local surface height.

Recently [6] we reported numerical results from two original models, LW2 and MM2, both resulting in B1-type step bunching. The equations of these models are constructed from terms identical in form but with presumably different values of the parameters in them. The procedure of model construction at this stage is to take the stabilizing (corresponding to step-step repulsion) term from a B2-type model and use it both as a stabilizing term in the new model and, with inverted sign and new symbols for the parameters in it, as a destabilizing (corresponding to step-step attraction) term. Thus LW2 is obtained applying this procedure to one of the most studied step bunching models we call LW after Liu and Weeks and MM2 is obtained from a minimal model called MM1 [6 and references therein]. These models, LW2 and MM2, remain in the B1-type for all studied values of model parameters and are shown to be distinguished only by the value of the exponent $\beta$ in the time-scaling of $N$ [7].

**The models.** We construct their equations for step velocity $dx_i/dt$ ($x_i$ is the current coordinate of the $i$-th step) from terms taken crosswise from the two *primary* models MM2 and LW2 to obtain:

$$\frac{dx_i}{dt} = -K\left[\left(\frac{l}{\Delta x_i}\right)^{p+1} - \left(\frac{l}{\Delta x_{i+1}}\right)^{p+1}\right] + U(2f_i - f_{i-1} - f_{i+1}) \quad \text{(MM2LW)}$$

and:

$$\frac{dx_i}{dt} = -K(2g_i - g_{i-1} - g_{i+1}) + U\left[\left(\frac{l}{\Delta x_i}\right)^{n+1} - \left(\frac{l}{\Delta x_{i+1}}\right)^{n+1}\right] \quad \text{(LW2MM)}$$

where

$$g_i = \left(\frac{l}{\Delta x_i}\right)^{p+1} - \left(\frac{l}{\Delta x_{i+1}}\right)^{p+1} \quad \text{and} \quad f_i = \left(\frac{l}{\Delta x_i}\right)^{n+1} - \left(\frac{l}{\Delta x_{i+1}}\right)^{n+1},$$

$K$ and $U$ are parameters with dimension of velocity, $l$ is the initial vicinal distance, and $\Delta x_i = x_i - x_{i-1}$ is the distance between the steps with number $i$ and $i$-1, called also *terrace width*. In the right hand side of the velocity equations first is the term that destabilizes the regular step train and second is the term that favors the equidistant step arrangement. It is interesting to note that the stabilizing term coming from LW2 is obtained considering the effect on the surface free energy caused by step-step repulsion with energy inversely proportional to the inter-step distance raised to power *n*. This term spans the widths of four terraces – the two adjacent to the step and the two next-nearest to it. The destabilizing term from LW2 is identical in form to the stabilizing one but with an opposite sign thus the exponent *p* is already exponent from a step-step attractions law. The magnitudes of the step-step interactions enter the parameters *K* and *U* which are with the dimension of velocity. The terms from MM2 contain only the widths of the two adjacent terraces and they are designed to mimic the realistic terms. Thus the parameters *K* and *U* in MM2 could not be linked back to clearly defined physical entities. Our nomenclature for the names of the models reflects their genealogy and it may look clumsy or even superfluous but we recall here the fact that there is one more candidate for a *primary* model, TE2 [10], of the same type as MM2 and LW2, obtained with proper generalization of the model of Tersoff et al. [4]. Thus, when using also terms from TE2 for model generation, the complete set of models consists already of six *hybrid* models. In such situation a suitable nomenclature is rather important.

**Numerical results.** In order to study the step bunching process in the models just introduced we integrate numerically their equations of step velocity using a fourth order Runge-Kutta routine. Usually 3000 steps are included in the calculations and periodic boundary conditions apply. In both models there is only step rearrangement without translation of the mass center of the step system that would reflect crystal growth or evaporation. Only LW2MM inherits from the *primary* models the B1-type of step bunching but in a rather complex fashion - the surface slope is largest in the both ends of the bunch (hence there appears the minimal inter-step distance in the bunch), then decreases inward and jumps again to a constant value in the inner part of the bunch, again independent of *N*, see Figure 1. MM2LW shows B2-type of step bunching with surface slope largest in the middle of the bunch [8] and decreasing symmetrically towards both ends of the bunch. Step trajectories in LW2MM are typical for B1-type with bunch coalescence and without exchange of steps between the bunches. In MM2LW we observe an interesting dynamic effect – some bunches increase to a certain size and then decrease emitting single steps towards the two neighboring bunches, Figure 2, a phenomenon similar to the Ostwald ripening.

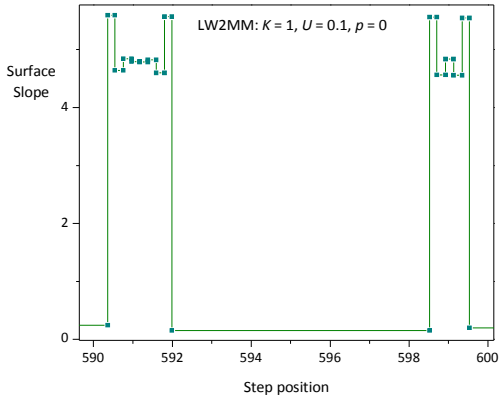 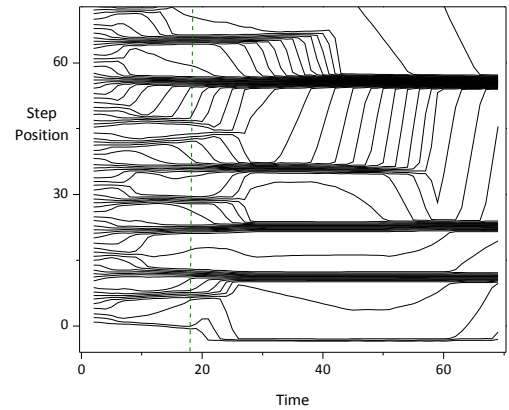

Figure 1. Surface slope from LW2MM, the higher slope corresponds to the bunches, their width is proportional to the number of steps in a bunch. The inter-step distance is inversely proportional to the slope.

Figure 2. Step trajectories from MM2LW show the phenomenon of bunch decay – a bunch with certain size starts to emit single steps towards the two neighboring bunches until complete disappearance.

Further we reveal some quantitative aspects of the bunching process. The statistics of bunches has in its basis an important quantity – the definition which step-step distance is defined as bunch one. We define it as any distance that is smaller than the initial (the vicinal) one. For the purposes of the quantitative analysis we employ two parallel monitoring schemes as described in [9] to obtain the size-scaling of the minimal inter-step distance in the bunch, $l_{min}$ vs. $N$, and time-scaling of the average number of steps in the bunch. Here we give only a brief account of this numerical procedure. In monitoring scheme I (MS-I) we calculate the average number of bunches with the time step of the numerical integration. If two adjacent distances are less than the vicinal each they belong to the same bunch. The number of bunches is used to find the average bunch width, average bunch distance, etc. In the second monitoring scheme (MS-II) are collected data during whole calculation separately for every bunch size that would appear at any time and at the end of the calculation the collected quantities such as minimal step-step distance, bunch width, etc., are averaged over the number of times the given bunch size appears.

Although the two schemes monitor different features of the bunching process they give concurrent results for the dependencies that can be obtained in both schemes as shown in [9] for the 'bunch width vs. bunch size' dependence. Here we show explicitly only the dependencies obtained for MM2LW, Figure 3 and Figure 4. As for LW2MM, it shows $n$-independent time-scaling of $N$ with universal exponent ~ 1/3, and $l_{min}$ independent of $N$.

**Conclusions.** Simple classification of the step bunching phenomena is proposed according to the behavior of the surface slope inside the step bunch with the increase of the number of steps $N$ in the bunch. It identifies two different types of step bunching – B1-type (slope is constant) and B2-type (slope increases). Two novel *hybrid* models of (potentially) unstable step motion are constructed and the bunching process is studied for a particular set of parameters. Interesting and complex behavior was observed that deserves further analysis varying especially *p*, the exponent in the destabilizing term in the equations for step velocity. Our results are expected to stimulate the search of adequate continuum equations. We extend the basis for understanding in detail the role of the different terms that enter the model equations and their hidden interplay. It is a matter of further intensive studies to identify properly the universality classes to which belong the two models.

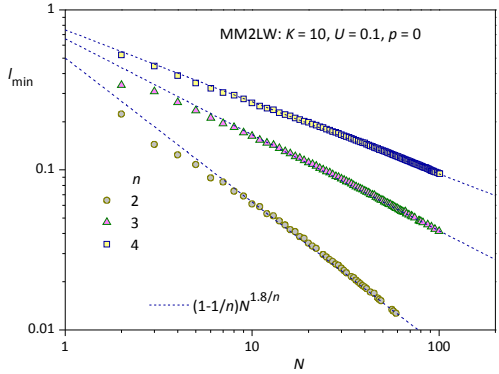 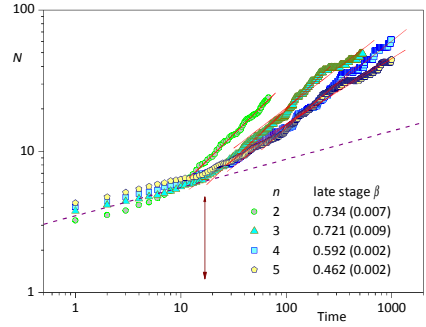

Figure 3. MM2LW, scaling of the minimal step-step distance $l_{min}$ in the bunch with bunch size $N$, monitoring scheme II (MS-II). $(K,U,p)=(10,0.1,0)$

Figure 4. MM2LW, time-scaling of the number of steps in the bunch $N$, monitoring scheme I (MS-I). $(K,U,p)=(1,0.1,0)$

**Acknowledgments.** The authors acknowledge support from the IRC-CoSiM and MADARA Grants from the Bulgarian National Science Fund.